\documentclass[conference]{IEEEtran}
\IEEEoverridecommandlockouts
\usepackage{cite}
\usepackage{amsmath,amssymb,amsfonts}
\usepackage{algorithmic}
\usepackage{graphicx}
\usepackage{gensymb}
\newcommand{\linebreakand}{%
  \end{@IEEEauthorhalign}
  \hfill\mbox{}\par
  \mbox{}\hfill\begin{@IEEEauthorhalign}
}
\usepackage{xcolor}
\def\BibTeX{{\rm B\kern-.05em{\sc i\kern-.025em b}\kern-.08em
    T\kern-.1667em\lower.7ex\hbox{E}\kern-.125emX}}
\begin{document}

\title{DEEPBEAS3D: Deep Learning and B-Spline Explicit Active Surfaces\\

}
\author{
\begin{tabular}{cccc}
   1\textsuperscript{st} Helena Williams & 2\textsuperscript{nd} João Pedrosa & 3\textsuperscript{nd} Muhammad Asad & 4\textsuperscript{th }Laura Cattani
   \\
    \textit{Department of} &
    \textit{INESC TEC \& Faculty} & 
    \textit{School of Biomedical} & \textit{Department of}
    \\
    \textit{Development \& } & 
    \textit{of Engineering} & 
    \textit{Engineering \& Imaging} & \textit{Development \&}
    \\
   \textit{Regeneration, KU Leuven} & \textit{University of Porto} & 
   \textit{Sciences, KCL} & 
   \textit{Regeneration, KU Leuven}
   \\
    Leuven, Belgium & 
    Porto, Portugal & 
    London, UK &
    Leuven, Belgium 
    \\
    0000-0002-7970-0290 &
    0000-0002-7588-8927 & 
    &
    0000-0001-5232-3619  
    \\
  \end{tabular}
\\
  \begin{tabular}{ccc}
    5\textsuperscript{th }Tom Vercauteren & 6\textsuperscript{th }Jan Deprest & 7\textsuperscript{th }Jan D'hooge 
    \\
    \textit{School of Biomedical } & \textit{Department of} & \textit{Department of } \\
    \textit{Engineering \& Imaging} & 
    \textit{ Development \&} & 
    \textit{Cardiovascular Sciences}
    \\
    \textit{Sciences, KCL} & 
    \textit{Regeneration, KU Leuven} &
    \textit{KU Leuven}
    \\
    London, UK & 
    Leuven, Belgium & 
    Leuven, Belgium 
    \\
     0000-0003-1794-0456 &
     &
\end{tabular}}

\maketitle

\begin{abstract}
Deep learning-based automatic segmentation methods have become state-of-the-art.
However, they are often not robust enough for direct clinical application, as domain shifts between training and testing data affect their performance.  
Failure in automatic segmentation can cause 
sub-optimal results that require correction. To address these problems, we propose a novel 3D extension of an interactive segmentation framework that represents a segmentation from a convolutional neural network (CNN) as a B-spline explicit active surface (BEAS). BEAS ensures segmentations are smooth in 3D space, increasing anatomical plausibility, while allowing the user to precisely edit the 3D surface. We apply this framework to the task of 3D segmentation of the anal sphincter complex (AS) from transperineal ultrasound (TPUS) images, and compare it to the clinical tool used in the pelvic floor disorder clinic (4D View VOCAL, GE Healthcare; Zipf, Austria). 
Experimental results show that: 1) the proposed framework gives the user explicit control of the surface contour; 
2) the perceived workload calculated via the NASA-TLX index was reduced by 30\% compared to VOCAL; and 3) it required 70\% (170 seconds) less user time than VOCAL
(p$<0.00001$).          
\end{abstract}


\section{Introduction}
Medical image segmentation is used for diagnosis, therapy monitoring, and surgical planning, however, manual segmentation requires clinical expertise and is labour-intensive, especially when providing consistent 3D segmentations beyond slice-based contours. Subsequently, there has been a drive for automatic segmentation with convolutional neural networks (CNNs) achieving state-of-the-art performance.
However, they often lack sufficient robustness and trustworthiness needed for clinical integration, as they can require corrections \cite{wang2018deepigeos}, mainly due to variations in image and acquisition quality. 
Interactive methods allow the improvement of automatic segmentation and clinical trust by utilising the user's anatomical knowledge and giving control to the end user \cite{luo2021mideepseg,wang2018deepigeos}.
 A two-stage approach was used in \cite{wang2018deepigeos}, where initial CNN segmentations were refined by scribble-based interactions. Real-time scribble-based feedback was achieved by utilising a lightweight CNN \cite{asad2022econet}. These demonstrated the value of utilising anatomical knowledge. However, scribble-based approaches lack explicit control of the surface making them less intuitive for end-users seeking fine grained control over the output. In \cite{williamsinteraction2d} explicit boundary interaction was achieved by a CNN and the B-spline explicit active surface (BEAS) \cite{beasbarboa2012} framework, allowing real-time, explicit control of the 2D contour. However, no 3D implementation of such approach has been demonstrated in previous work, and the work in \cite{williamsinteraction2d} requires manually tuning of BEAS parameters. 
Motivated by this, we propose a 3D interactive segmentation tool based on 3D deep learning and BEAS (DeepBEAS3D), expanding on the previous 2D method \cite{williamsinteraction2d}. 
DeepBEAS3D allows real-time explicit control and interaction of the surface, and benefits from BEAS's inherent regularising effect, meaning no segmentation post-processing is needed.
DeepBEAS3D is compared against Virtual Organ Computer-Aided Analysis (VOCAL) (GE Healthcare; Zipf, Austria), a clinical gold standard tool for 3D US segmentation.
Our experimental validation uses 30 3D TPUS of the anal sphincter, comparing the time and perceived workload required (measured using the NASA TLX index \cite{pandian2020nasatlx}) to segment the anal sphincter complex (AS) to a clinically acceptable standard. The AS consists of a group of muscles that surround the anal canal, ensuring continence and regulating bowel movements. While AS volumetric segmentation lacks current clinical use, its clinical significance lies in the assessment of anal sphincter injury, a debilitating condition associated with obstetric anal sphincter injuries. Precise AS location and orientation in the TPUS, and pathology, such as intact muscle status, plays a vital role in assessment. 
Our contributions are a novel 3D extension of a CNN and BEAS-based interactive segmentation framework and a novel application of hyper-parameter optimisation for wider applicability of DeepBEAS3D and BEAS. 
\section{Material and Methods}
\subsection{Methodology}
\label{sec:methods}
\begin{figure*}[btp]
    \centering
    \includegraphics[width=0.75\textwidth]{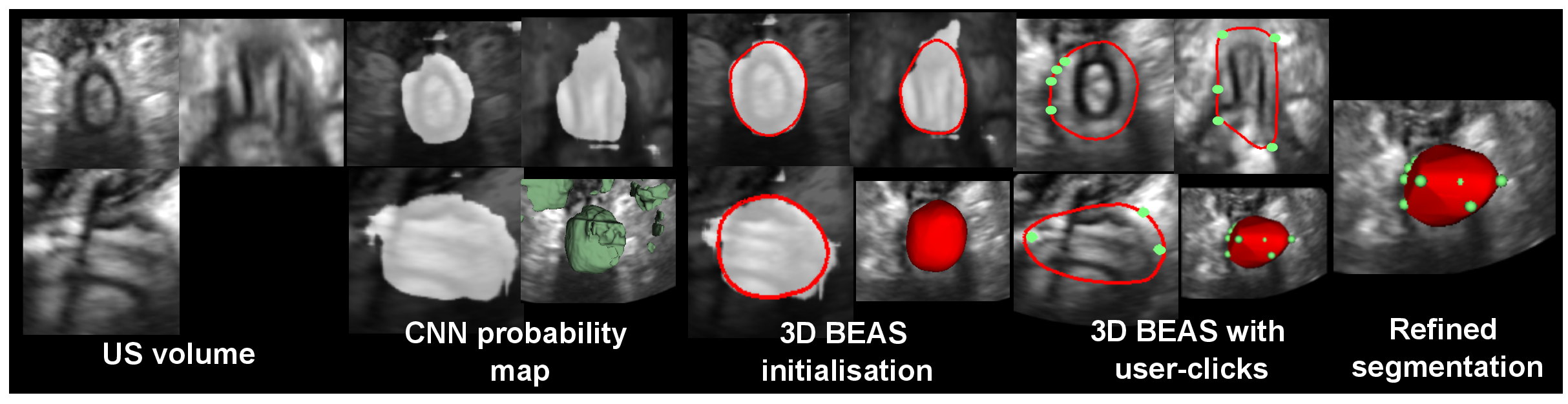}
    \caption{DeepBEAS3D shows an input 3D TPUS volume, initial probability map from CNN model (white) and the 3D rendered probability map ({\color{green}green surface}), 3D BEAS initialisation ({\color{red} red outline}), refined BEAS using user interaction ({\color{green} green points}) and the BEAS surface ({\color{red} red surface}).} \label{pipelinev2}
\end{figure*}
 Fig.~\ref{pipelinev2} shows the pipeline consisting of a 3D U-Net that segments the AS; BEAS evolves to represent the output 3D probability map as an active surface, and in real-time the interactive energy adapts the surface using user-defined boundary clicks.  

\subsubsection{BEAS evolution}
The CNN probability map is represented as a BEAS surface, where BEAS represents a coordinate of the surface explicitly as a function of the remaining coordinates. Here, BEAS is used to model a complete 3D object in the spherical domain with a defined origin, a good approximation for the AS, which has a convex shape. We represent the surface radius from the origin as an explicit function of both azimuthal and zenithal angles ($i.e., \rho = \psi(\theta,\varphi)$).
$\rho$ can be defined in n-dimensional space, as a linear combination of $n-1$ dimensional B-spline basis functions of degree, $d$, $\beta^d_h(\cdot)$ \cite{beasbarboa2012,SPLINES,4895341}.
The mesh points (i.e., knots) of the B-splines are located on a grid defined in the spherical coordinate system, with regular spacing given by $h$, where $h=2^{s}$ and $s$ is the scale.
The number of knots, $N_k$, depends on the azimuthal mesh size, ${N}_{\theta}$, and zenithal mesh size, ${N}_{\varphi}$ ($N_k= {N}_{\theta}*{N}_{\varphi}$). BEAS is initialised as a sphere with a fixed origin and radius from the centre of mass and average radius of the probability map, respectively. 
The minimisation of an energy criterion, $E_{CNN}$, evolves the surface boundary towards the CNN probability map boundary.
$E_{CNN}$ is the localised Yezzi energy \cite{localyezzi,Yezzi2002AFG}, which maximises the localised difference of voxel intensity between the contours interior and exterior. 
For more detail, we refer the reader to \cite{beasbarboa2012,williamsinteraction2d}.
\subsubsection{Interaction}
\begin{figure}[t]
    \centering
    \includegraphics[trim=1.0cm 0.3cm 1.0cm 0.3cm,width=0.44\textwidth]{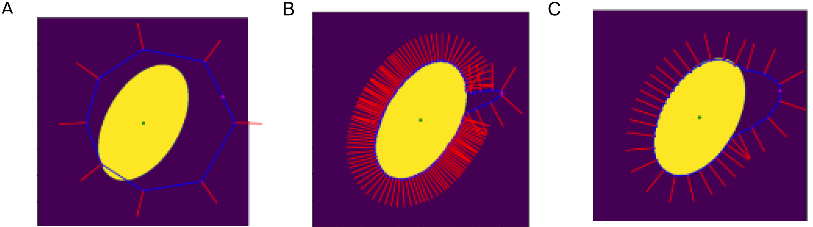}
    \caption{Number of knots vs. the ease of user interaction, a) too few knots showing large radial movement, b) too many knots showing \emph{spiky} behaviour, c) Goldilocks zone: a trade-off between radial movement and spiky behaviour.} 
    \label{beasdiff}
\end{figure}

The interactive energy, $E_i$, is driven by the parametric radial distance of the user point, $\rho_u$, and the closest BEAS position, $\rho_m$, (i.e., $D=(\rho_u-\rho_m)^2$). Here, we refer to the explicit function $\rho=\psi(\theta,\varphi)=\psi(\mathbf{x}^*)$ for brevity. $E_i$ is minimised w.r.t. each B-spline coefficient $c[k_i]$ as:
\begin{equation}
    \frac{\partial E_i}{\partial c[k_i]} = 2 \int \!\!\!\int_{\Gamma} \delta(\mathbf{x}^* - \mathbf{x}^*_u)(\psi(\mathbf{x}^*)-\rho_u)\beta^d_h \left( \mathbf{x}^*-hk_i\right) d\mathbf{x}^*,
\end{equation} where $\Gamma$ is the closed surface, $\mathbf{x}^*_u$ is the user point in spherical coordinates \cite{interactivebeas}.  
$\delta(\mathbf{x}^* - \mathbf{x}^*_u)$ is the Dirac delta function, which is non-zero at the position $\mathbf{x}^* = \mathbf{x}^*_u = (\theta_{u},\varphi_{u})$. 
The total energy, $E_{total}$, driving BEAS on interaction is compounded of three terms: $E_U$ (Localised Yezzi energy of the TPUS), $E_{CNN}$, and $E_{i}$:
\begin{equation} \label{eq:4}
    E_{total}= \alpha E_{U} + \eta E_{CNN} + \gamma E_{i},
\end{equation}
where $\alpha$, $\eta$ and $\gamma$ are hyper-parameters.
Fig.~\ref{beasdiff} illustrates the ease of user interaction of BEAS in relation to mesh sizes. 
\subsubsection{BEAS hyper-parameter tuning}
BEAS hyper-parameters that require fine-tuning include the mesh sizes, $N_{\theta}$ and $N_{\varphi}$ and scale, $s$. These affect the smoothness and the ease of updating from user interaction (Fig.~\ref{beasdiff}).
This automated hyper-parameter tuning protocol defines an optimisation energy term, E, to ensure that BEAS has suitable smoothing and is intuitive regarding user interaction updates, defined here as the \emph{Goldilocks zone}. Within the \emph{Goldilocks zone} there is a trade-off between foreground (i.e., surface close to the point of interaction) curvature movement and background (i.e., surface far from the point of interaction) radial movement.
Optimisation energy, $E$, is defined and it's minimum value determines the BEAS mesh and scale that correspond to the \emph{Goldilocks zone}:
 \begin{equation} \label{eq:search}
     E = \frac{1}{DSC(I_{C},I_{S})} + HD(I_{C},I_{S}) + 2 (K_{S}-K_{I}) + (r_{S}-r_{I}),
 \end{equation}
where $DSC$ and $HD$ are the Dice and Hausdorff distance between the CNN segmentation, $C$, and the BEAS surface, $S$;
$K$ and $r$ are the local Gaussian curvature and radial distance \cite{Pressley2010}, and $I$ is the BEAS surface after interaction.
$K$ has a weighting factor of 2, to balance the effect of DSC and HD where a larger mesh size results in a lower HD and higher DSC score, favouring higher mesh sizes. 

To measure the ease of interaction, five randomly simulated user points are added in spherical space within a radius of $(10\%- 20\%)$ from the initial surface. After adding each point, the background radial and the foreground local Gaussian curvature change are measured. After all points are added, E is measured. The hyper-parameter tuning is first applied to one AS training label in a brute force search, with a large scale and mesh size range (i.e., $N_{\theta}$ $[6...24]$, $N_\varphi$ $[6...24]$ and scales $[0,1]$). The global minimum energy observed for this training label is at a mesh size of $[12,16]$ and scale of $0$. Then, tuning is repeated on four new training labels in a refined search, with a set scale and smaller mesh size range (i.e, $N_{\theta}$ $[12...24]$, $N_\varphi$ $[12...24]$ and a scale of 0). The refined range is defined by the minimum and maximum mesh size that achieved an error less than 10\% higher than the global minimum energy at $[12,16]$. This identifies several combinations of optimal parameters ($[12,16], [20,16],[16,20],[20,16]$,$[16,20]$), where $[12,16]$ is chosen as it has the smallest mesh size. These parameters are assumed to be suitable for all AS segmentations, as in literature they are not adjusted from image to image \cite{pedrosabeas2017}. 
\subsection{Data collection and experimental details}
\subsubsection{Data collection} Patients gave informed consent for secondary analysis of images for research purposes (S63009) and analysis of anonymised TPUS volumes was retrospective. The CNN was trained and tested on a dataset of 115 and 30 TPUS volumes, respectively. 
AS ground truth labels of the training dataset were acquired by expert L.C. (with over five years' experience in AS TPUS imaging and four years' experience in VOCAL). All volumes were acquired at UZ Leuven, Belgium, following the clinical protocol defined by Dietz et al. (i.e., transperineal transducer placed on the perineum in the coronal plane) \cite{DIETZAS} on a Voluson E10 US system (GE Healthcare). 
\subsubsection{VOCAL protocol}
VOCAL segments by rotating the structure around a fixed contour axis and 2D contours are manually delineated in each plane. We defined the axis along the AS central axis and a rotation angle of $30$\textdegree{} in a $180$\textdegree{} range was used. Therefore, after six manual delineations a complete 3D segmentation was formed, as VOCAL assumed a symmetrical structure. 
VOCAL then allows manual contour corrections in other surface planes \cite{Gontard2021}.  
\subsubsection{3D Slicer protocol}
DeepBEAS3D was implemented into a 3D Slicer module (code is planned to be released upon publication). For segmentation, the `Run BEAS' button is selected and an initial BEAS surface displays over the TPUS. The user visually assessed the segmentation and added boundary points. The protocol used was to assess the axial and midsagittal planes first, to ensure the total AS length and width were included. Finally, refinement in the coronal plane would be made if necessary. 
\subsubsection{Implementation details}
DeepBEAS3D was implemented on a 24GB NVIDIA Quadro P6000 on a Windows desktop. 
3D U-Net \cite{unet3D} was implemented using MONAI~\cite{MONAI2020} and training was performed using a Dice loss and a Nesterov optimiser (batch size=3, momentum=0.99, learning rate=0.001, learning rate decay=0.9 and weight decay=$3\cdot10^{-5}$) for 1000 epochs. 100, 15 and 30 TPUS volumes were used for training, validation and testing respectively.  
Data augmentation included random scaling (-30\%, +40\%), x-axis rotation (-45\textdegree, +45\textdegree) and Gaussian noise. Validation of the network training was performed every 100 epochs and early stopping based on the highest validation Dice score was employed, to select the model at epoch 300 for inference. 
BEAS was run on an Intel(R) Core(TM) i9-7900X CPU @ 3.30GHz, the neighbourhood size used to estimate the probability map was 100 voxels ($\approx 30mm$), allowing recovery from a poor initialisation. Optimisation determined BEAS to be discretised into [12,16] knots, with spacing, $h$, set to $1$. 
For interaction, in \eqref{eq:4} $\alpha = 1$, $\eta= 0.3$ and $\gamma = 1$ and the neighbourhood size was 10 voxels ($\approx3mm$) based on  \cite{williamsinteraction2d}.
\section{Results and discussion} 
\begin{figure}[t!]
\centering
\includegraphics[trim=0.8cm 0.3cm 0.8cm 0.3cm,width=0.5\textwidth]{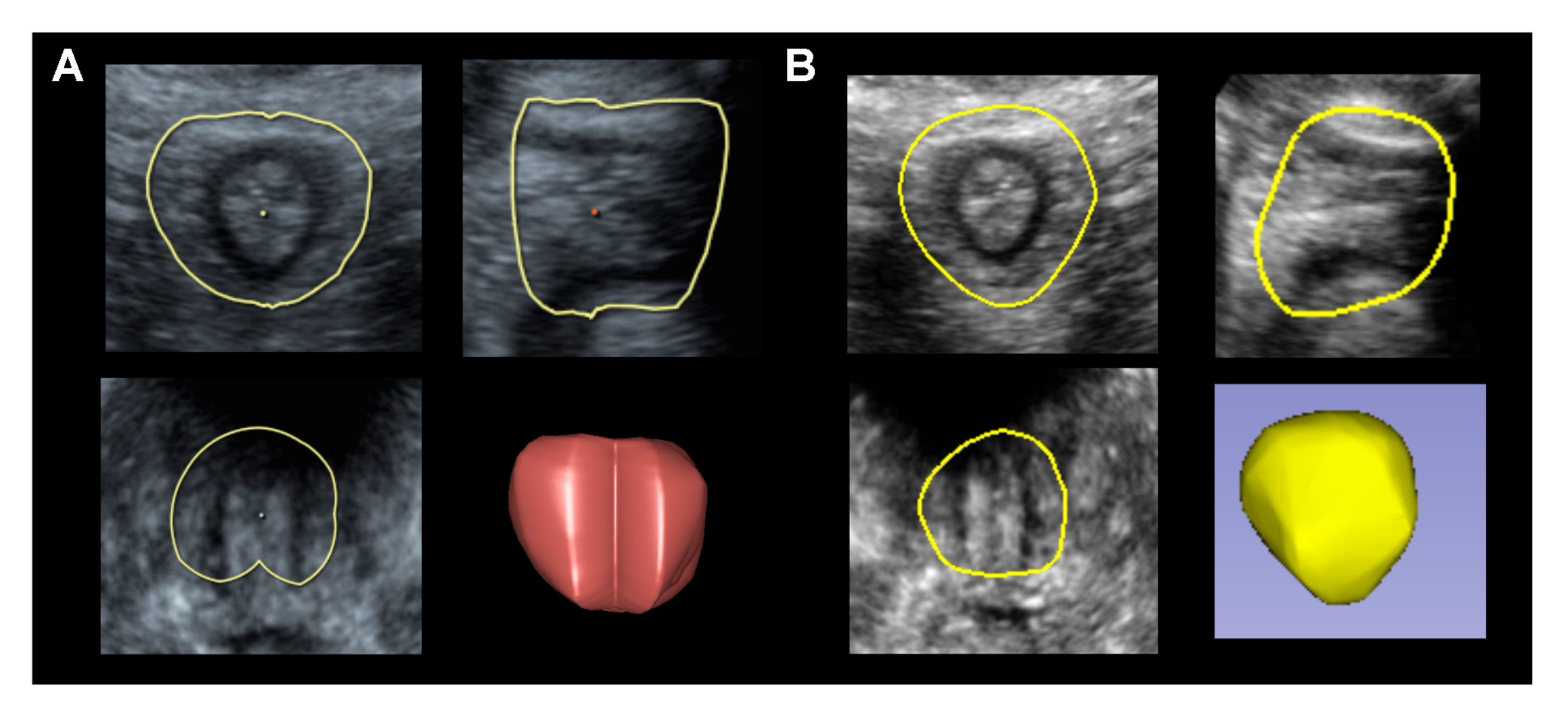}
\caption{Visualisation of segmentations obtained with A) VOCAL and B) DeepBEAS3D showing coronal (top left), midsagittal (top right) and axial (bottom left) planes, and 3D reconstruction (bottom right). 
}
\label{fig:images}
\end{figure}

Fig.~\ref{fig:images} shows a segmentation using VOCAL and DeepBEAS3D which are visually `similar'. 
The clinician agreed DeepBEAS3D accomplished a clinically acceptable standard for all volumes and achieved a `clinical acceptability' of $100\%$. However, only 2 CNN and CNN + BEAS initialised (i.e., DeepBEAS3D without user interaction) segmentations required no editing and achieved a `clinical acceptability' of $7\%$. 
VOCAL curations on average took the clinician $247 \pm 74$ seconds, whereas DeepBEAS3D took $76 \pm 44$ seconds. 
DeepBEAS3D time does not include CNN inference, to be independent of CNN performance. Here, the average inference time was $12$ seconds and DeepBEAS3D's total time was significantly lower (paired t-test with a significance level of 0.05, p$<0.00001$) than VOCAL. 
Table~\ref{tab:nasa_tlx} shows the mean NASA-TLX scores for segmentation \cite{pandian2020nasatlx}. 
\begin{table}[t]
\label{tab:nasa_tlx}
\centering
\caption{The perceived weighted workload score and sub-scale scores for VOCAL and DeepBEAS3D. A low score corresponds to less effort, less frustration, less mental, temporal and physical demands, and a higher performance.}
\begin{tabular}{c|cc}
\hline
NASA-TLX        & \multicolumn{2}{c}{Average}   \\ 
weighted scores &VOCAL & DeepBEAS3D  \\ \hline
Effort          & 16.00 & 10.00   \\
Frustration     & 12.00 &  10.00 \\
Mental Demand   & 6.67 & 6.67      \\
Performance     & 6.67 & 3.33   \\
Physical Demand & 1.33  &  0.00     \\
Temporal Demand & 0.00  & 0.00      \\ \hline
Total workload  & 42.67 & 30.00  \\ \hline
\end{tabular}
\end{table}
In Fig.~\ref{fig:images}, the top right image shows a discrepancy due to intra-observer variability. The other visual difference is due to the smoothing nature of BEAS, which creates a biologically plausible structure (i.e., DeepBEAS3D does not include sharp features shown in the bottom left image).
Sub-optimal CNN segmentations were due to poor TPUS acquisition, AS tearing, US shadowing (shown in Fig.~\ref{fig:images}), which reduced the visibility of the AS.
The proposed BEAS optimisation algorithm was applied to five labels, which identified a range of hyper-parameters and the smallest mesh size was chosen to reduce the complexity of mesh representation. 
In situations demanding the editing of finer detailed structures, either involving a larger mesh size or aligning with the specific CNN output, tailored hyper-parameters can be selected from the range identified in Section~\ref{sec:methods}.

DeepBEAS3D and BEAS improved the CNN segmentation by removing multiple components of smaller volume, smoothing the surface, and removing non-anatomical holes, solving common issues of voxel-wise segmentation. 
DeepBEAS3D required 70\% less user time than VOCAL, and 
user time may further reduce with increased user exposure. As we have seen in our study, the time to delineate using DeepBEAS3D decreased from 94 seconds for the first 15 TPUS to 60 seconds for the last 15. 
Table~\ref{tab:nasa_tlx} shows that the expert perceived DeepBEAS3D to have a lower workload than VOCAL. 
In particular, DeepBEAS3D reduced the perceived effort, possibly because, in most cases, an adequate initial segmentation is presented. Therefore, only slight editing is required, whereas VOCAL requires several manual 2D delineations and editing.

The main advantage of DeepBEAS3D to other interactive segmentation 
methods is that it gives the user explicit control of the surface contour, without relying on CNN refinement, meaning specific points can be defined as the surface boundary and updates are applied in an efficient CPU-based implementation. This feature is significant in the context of medical device regulation and liability, as well as establishing trust between clinicians and automated segmentation tools. This could be utilised by other 3D segmentation tasks once the 3D Slicer module is publicly available (upon publication). The hyper-parameter tuning methodology allows for wider adoption, as it does not require manual optimisation. This is the first work to apply such an algorithm for tuning BEAS parameters for the ease of interaction and segmentation accuracy. We believe it will increase the applicability and usability of BEAS and DeepBEAS3D to other medical imaging problems. 
 
The selected BEAS mesh hyper-parameters favoured little background radial movement while reducing spiky behaviour close to the point of interaction, which can be common when interacting with splines.
DeepBEAS3D segmentations had better anatomical plausibility as they inferred consistently across all volume slices. In future work, it could be beneficial to extend DeepBEAS3D to work jointly for several structures and modalities. It would also be beneficial to refine the proposed BEAS hyper-parameter tuning algorithm by calculating the combined average energy of several AS training labels to avoid under or over-fitting.
We also aim to use hyper-parameter optimisation to determine hyper-parameters for equation \eqref{eq:4}. 
\section{Conclusion}
In this work, a novel 3D interactive segmentation method, DeepBEAS3D, was proposed. DeepBEAS3D guides the segmentation through the radial distance between the user-defined point and the mesh surface and gives the user explicit control of the surface contour, without relying on a CNN for refinement. 
DeepBEAS3D requires less user-time and has a lower perceived workload than VOCAL, a clinical tool currently used in Obstetrics and Gynaecology clinics, suggesting it could benefit the clinical workflow. BEAS ensured the segmentation was a single component of maximum volume, smooth and complete. Hence, it was more anatomically plausible than the CNN segmentation alone.
\bibliographystyle{ieeetr}
\bibliography{DeepBEAS3D}

\end{document}